\documentclass[aps,amsmath,amssymb,prb,reprint,showpacs]{revtex4-1}
\usepackage[toc, page]{appendix}
\usepackage{graphicx}
\usepackage{tikz}
\usepackage{tikzscale}
\usepackage{float}
\usepackage{braket}
\usetikzlibrary{shapes,arrows,positioning,patterns,calc,decorations.markings}

\newcommand{\diff}[1]{\mathrm{\frac{d}{d\mathit{#1}}}}
\newcommand{\op}[1]{\mathrm{#1}}

\newcommand{\bs}{\boldsymbol}
\newcommand{\ua}{\uparrow}
\newcommand{\da}{\downarrow}

\newcommand{\bfa}{\mathbf{a}}
\newcommand{\bfb}{\mathbf{b}}
\newcommand{\bfg}{\mathbf{g}}

\newcommand{\bfT}{\mathbf{T}}

\newcommand{\bfr}{\mathbf{r}}
\newcommand{\bft}{\mathbf{t}}

\newcommand{\bfv}{\mathbf{v}}

\newcommand{\bfI}{\mathbf{I}}

\newcommand{\bfk}{\mathbf{k}}

\newcommand{\bfalp}{\boldsymbol{\alpha}}
\newcommand{\bfbet}{\boldsymbol{\beta}}

\newcommand{\bftau}{\boldsymbol{\tau}}
\newcommand{\bfsig}{\boldsymbol{\sigma}}

\newcommand{\tr}{\text{tr}}
\newcommand{\orbA}{\overrightarrow{\mathbf{a}}}
\newcommand{\orbB}{\overrightarrow{\mathbf{b}}}

\newcommand{\orj}{\overrightarrow{j}}
\newcommand{\orbalp}{\overrightarrow{\bfalp}}
\newcommand{\orbbet}{\overrightarrow{\bfbet}}

\newcommand{\olj}{\overleftarrow{j}}

\newcommand{\kpar}{\bfk_{\parallel}}
\newcommand{\eferm}{E_{\text{F}}}
\newcommand{\bfone}{\mathbf{1}}
\newcommand{\bfzer}{\mathbf{0}}
\newcommand{\bfrprime}{\mathbf{r'}}
\newcommand{\bftprime}{\mathbf{t'}}

\newcommand{\Figref}[1]{Fig.~\ref{#1}}

\newcommand{\figscale}[4]{%
\begin{figure}%
	\begin{center}%
		\includegraphics[scale=#1]{#2}%
	\end{center}%
	\caption{\footnotesize #3}%
	\label{fig:#4}%
\end{figure}%
}

\begin{document}

\title{Application of the Landauer-B\"uttiker method to the calculation of interlayer exchange coupling in closed ballistic multi-layers}

\author{V. Fadeev, A. Umerski}
\affiliation{Department of Mathematics and Statistics, Open University, Milton Keynes MK7 6AA,
U.K.}

\date{\today}

\begin{abstract}
In this communication we study the behaviour of spin current components in a ballistic junction consisting of two semi-infinite leads and a scattering region composed of two magnetic layers separated by a non-magnetic metallic spacer. We then consider the system being gradually isolated from the leads, which we refer to as the transition from an open to a closed regime.
As expected on physical grounds, charge and in-plane components vanish, but the out-of-plane spin current remains finite and gives rise to the oscillatory interlayer exchange coupling (IEC) between the magnets.
We show that the out-of-plane spin current reduces to a set of peaks in momentum space, which correspond to discrete energy eigenstates of the closed system.
Furthermore, we demonstrate that the expression for the IEC in terms of spin current for an open system reduces to an expression in the closed system, which corresponds exactly to the definition of IEC in terms of energy difference.
\end{abstract}

\pacs{75.76+j, 72.25.Ba, 73.63.-b, 73.40.-c, 73.50.-h, 75.30.Et}

\maketitle

\section{Background}
The Landauer-B\"uttiker method is usually employed to calculate ballistic coherent transport properties of electron charge through materials subject to an electrical bias \cite{5392683}. An important feature of the method is that it is applied to \textit{open} systems consisting of a central scattering region connected to macroscopic reservoirs (terminals) via reflectionless leads. Carriers are emitted at all attainable energies, momenta and spin orientations and eventually escape back into the reservoirs where they undergo phase randomisation and do not contribute to transport again.

The Landauer formula \cite{PhysRevB.23.6851, Stone:1988:MYM:49387.49393} relates the conductance $G$ of the sample to the probability of electrons transmitting through it via the available energy levels. In the two-terminal case it can be stated as follows
\begin{equation}
G=\frac{e^2}{h}\op{tr}{\{\bft\bft^{\dagger}\}}, \label{eq:linresponse1}
\end{equation}
where transmission between the incoming and the outgoing modes is described by an $M\times M$ matrix $\bft$, $M$ being the number of modes and each component $t_{ij}$ denotes the transmission amplitude between modes $i$ and $j$. In particular, the modes can represent spin bands in the case of spin-polarized transport. Components of $\bft$ can be extracted from the scattering matrix, however, in this discussion we will find it more convenient to use the language of the \textit{transfer} matrix instead.

The Landauer-B\"uttiker formalism has been used to study components of spin current \cite{UmerskiSC} arising in a multi-layer system consisting of two magnetic layers separated by a non-magnetic spacer. Extensions to multi-terminal geometries have also been considered \cite{PhysRevB.78.085301}. It was also used \cite{PhysRevLett.84.2481} in the development of circuit theory of non-collinear multi-layers, including the cases of both diffuse, and ballistic contacts. A calculation of the Gilbert damping constant in \cite{PhysRevLett.88.117601} is similarly performed in terms of the transmission and reflection amplitudes.
The assumption common to all the referenced models is that the system is open, that is, the scattering region is connected to macroscopic reservoirs characterised by the respective chemical potentials.
There exist phenomena of interest that manifest themselves in the absence of any flow of carriers to or from the reservoirs. One such example is given by the interlayer exchange coupling (IEC) \cite{PhysRevB.52.411} defined as the interaction between two ferromagnets (FM) via itinerant electrons in a non-magnetic metallic spacer (NM). IEC is characterised by the difference in energy between the configurations where magnetisations of the FM layers are in the parallel (P) or anti-parallel (AP) alignment. It is shown \cite{Edwards:2007} that IEC can equivalently be calculated by accounting for the torque exerted by the total out-of-plane component of spin current incident on the NM/FM interface. Since the effect can exist in a closed system, it is not immediately obvious whether a direct application of the Landauer-B\"uttiker method and the transfer matrix is going to lead to meaningful conclusions. A more careful inspection suggests that the principal difference between the closed and the open setting is contained in the boundary conditions. Whereas in the open systems those typically represent unit waves arriving from the leads into the scattering region, it is clear that in a closed model we require a different set of compatible conditions to account for the successive reflections that electrons experience in the multi-layer. We impose such boundary conditions and proceed to verify their validity as follows. We derive expressions for IEC using two methods referred to above, torque and energy difference, and demonstrate that they lead to the same result. We expand the obtained expression into a sum over the energy eigenstates and show that the same set of energy values is obtained directly from the boundary conditions for the closed system.

We then proceed to isolate the system by introducing additional potential barriers and gradually increasing their height. Since the analytical treatment of the resulting multi-layer would not be particularly illuminating, this part of the demonstration is performed numerically. We show that the charge current and the in-plane spin current vanish, but the out-of-plane spin current remains finite.
As the transition occurs, the distribution of the out-of-plane spin current density in momentum space reduces to a set of delta-like resonant peaks. Furthermore, positions of these resonances coincide with the positions of the discrete set of energy eigenstates of the closed system.
We show analytically that the expression for the IEC in the Landauer formalism in terms of spin current for an open system reduces to an expression in the closed system which corresponds exactly to the definition of the IEC in terms of energy differences. The outcomes of this investigation give a physically appealing picture of the behaviour of spin current and the IEC as we move from open to closed systems. It also provides an interesting study of the behaviour of the Landauer formalism, applied to spin dependent phenomena, and demonstrates that with appropriate boundary conditions the Landauer formalism is applicable to both systems. 

This communication is organised as follows. In Section \ref{sec:tm} we summarise the necessary elements of the transfer matrix formalism and state the expression of spin current in terms of reflection matrices. We then use this expression in Section \ref{sec:iec} to write down the expression for IEC. In Section \ref{sec:bc} we derive boundary conditions for the closed system. We show that the compatibility requirement for those conditions gives rise to an eigenvalue equation, and that the solutions of that equation are equivalently obtained by expanding the IEC using the residue theorem. In Section \ref{sec:num} we give a numeric illustration of the transition to the closed regime and discuss how each of the current components behaves under the circumstances.

This work contains material from Chapter 5 of the PhD thesis \cite{oro71614} written by V. Fadeev under the supervision of A. Umerski.

\figscale{0.6}{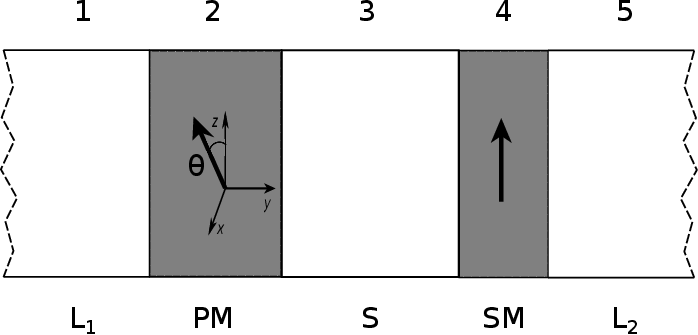}{Schematic of the multilayer structure consisting of a non-magnetic spacer (S), sandwiched between polarising (PM) and switching (SM) magnets and connected to semi-infinite non-magnetic leads $\text{L}_1$, $\text{L}_2$}{cpp}.

\section{Transfer Matrix Method and spin current}\label{sec:tm}
We consider a current-perpendicular-to-plane (CPP) geometry grown in the $y$ direction, consisting of two finite magnetic layers (a polarizing magnet $\text{PM}$ and switching magnet $\text{SM}$) separated by a non-magnetic conducting spacer $\text{S}$ and connected to semi-infinite non-magnetic leads ($\text{L}_1$ and $\text{L}_2$, respectively) (see Fig.~\ref{fig:cpp}) 
We choose the spin quantisation axis to be in the $z$ direction and take the magnetisation of the $\text{PM}$ to be rotated in the $xz$ plane by angle $\theta$. The majority and minority carrier potentials in the magnets are displaced by the exchange splitting energy.  
The entire system is assumed to have full rotational symmetry in the $xz$ plane, and transport is taken to be to be fully phase-coherent.

We are interested in calculating the wavefunction in the spacer due to electrons incident from the left lead. In a non-magnetic layer with potential $V$ the wavefunction is given by
\begin{equation}
\bs{\psi}=
e^{i k y} \bfalp +
e^{-i k y} \bfbet,
\label{eq:wf1}
\end{equation}
where $\bfalp=\begin{bmatrix}
\alpha^{\uparrow}& \alpha^{\downarrow}
\end{bmatrix}^T$,  
$\bfbet=\begin{bmatrix}
\beta^{\uparrow}& \beta^{\downarrow} 
\end{bmatrix}^T$ and $k = \sqrt{\frac{2m}{\hbar^2}\left(E-V\right)-k_x^2-k_z^2}$ is the out-of-plane component of the wave vector in the layer.
We are interested in the total transmitted amplitude in the spacer contributed by carriers arriving from the leads. Transmission and reflection amplitudes to the left (right) of an interface are determined by the action of $2\times 2$ matrices $\bft$, $\bfr$ ($\bft'$, $\bfr'$) respectively, on the incident wave, as shown in Fig.~\ref{fig:transmission}. We index the layers from 1 to 5, and write $\bft_{mn}$ ($\bfr_{mn}$) for transmission (reflection) from layer $m$ to layer $n$. Transfer matrix $\bfT_{mn}$ relates the amplitudes on the left to ones on the right, and can be shown to have the following structure
\begin{equation}
\bfT_{mn}=\begin{bmatrix}
\bftprime^{-1}_{mn} & -\bftprime^{-1}_{mn}\bfrprime_{mn}\\[10pt]
\bfr_{mn}\bftprime^{-1}_{mn} & \bft_{mn}-\bfr_{mn}\bftprime^{-1}_{mn}\bfrprime_{mn}
\end{bmatrix}. \label{eq:tm_struct}
\end{equation}
\figscale{0.6}{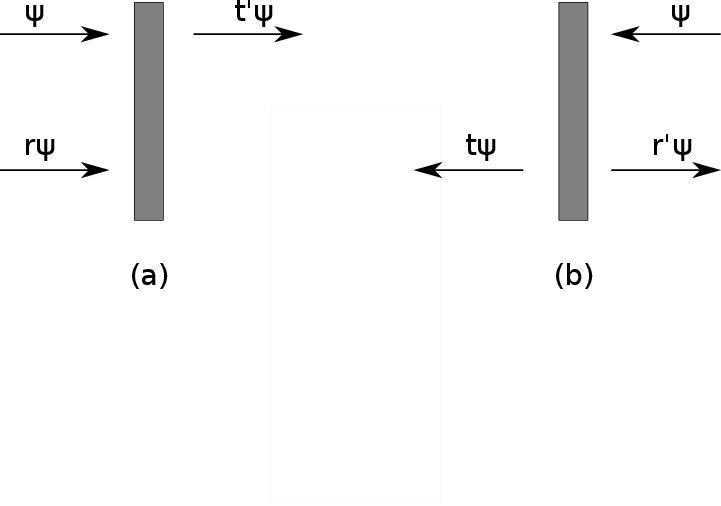}{Transmission and reflection amplitudes for the state of unit amplitude incident on the left (a) and the right (b) of the scattering interface.}{transmission}
If the incident electrons in $\text{L}_1$ have amplitude $\orbalp_1$, then the right- and left- amplitudes, $\orbalp_3$ and $\orbbet_3$, respectively, are given by
\begin{equation}
	\orbalp_3=\orbA_3\orbalp_1, \quad \orbbet_3=\orbB_3\orbalp_1, \label{eq:amp-spacer}
\end{equation}
where $\orbA_{3}$ and $\orbB_{3}$ account for the total right- and left-moving amplitude in the spacer, including all possible reflections within the layers. These can be calculated by summing over Feynman paths in the ladder approximation, or by using the semigroup property of the transfer matrix
\begin{equation*}
\bfT_{15} = \bfT_{13}\bfT_{35},
\end{equation*}
giving the following expressions
\begin{equation}
        \orbA_{3} =\left(\bfone-\bfrprime_{13}\bfr_{35}\right)^{-1}\bft'_{13}, \quad
        \orbB_{3} =\bfr_{35}\orbA_3,
    \label{eq:orab}
\end{equation}
where $\bfone$ is the $2\times 2$ identity matrix.
If we define $\bs{\sigma}_{0}=(2e/\hbar) \bfone$ then the charge and spin current components $j_{\nu}$, $\nu=0,x,y,z$, are given in terms of the wavefunction by
\begin{equation*}
    j_{\nu}=\frac{\hbar^{2}}{4mi}\left(\bs{\psi}^{\dagger}\bs{\sigma}_{\nu}\bs{\psi}'-{\bs{\psi}'}^{\dagger}\bs{\sigma}_{\nu}\bs{\psi}\right),
\end{equation*}
where $\bfsig_{x},\bfsig_{y}$ and $\bfsig_{z}$ are the Pauli matrices.  
Spin current in the spacer, due to an electron incident from the left-hand lead with amplitude $\orbalp_1$, is then given by
\begin{equation}
    \orj_{\nu}=k\left(\orbalp_1^{\dagger}\orbA_3^{\dagger}\bfsig_{\nu}\orbA_3\orbalp_1-
    \orbalp_1^{\dagger}\orbB_3^{\dagger}\bfsig_{\nu}\orbB_3\orbalp_1\right), \label{eq:current-spacer}
\end{equation}
where $k$ is the out-of-plane wave vector component in the spacer. By assumption, the electrons arrive in the leads with both spin orientations. Therefore, we evaluate \eqref{eq:current-spacer} under two boundary conditions, $\orbalp_1=\begin{bmatrix} 1 & 0 \end{bmatrix}^T$ (spin-up) and $\orbalp_1=\begin{bmatrix} 0 & 1 \end{bmatrix}^T$ (spin-down). The result, after summing over the contributions from both spin orientations, can be expressed in terms of a trace
\begin{equation}
\orj_{\nu}=k\,\tr\left(\orbA_3^{\dagger}\bfsig_{\nu}\orbA_3-
\orbB_3^{\dagger}\bfsig_{\nu}\orbB_3\right). \label{eq:current-trace}
\end{equation}
Calculation for the left-moving current density $\olj_{\nu}$ proceeds in a similar way.
In Appendix A it is shown that the out-of-plane component of the current density calculated using this method can be expressed in terms of reflection matrices only, and happens to be an exact derivative with respect to the polarization angle
\begin{equation}
\begin{aligned}
\overrightarrow{j}_{y}+\overleftarrow{j}_{y} & = -4k \Im\frac{\mathrm{d}}{\mathrm{d}\theta}\ln\det{\left(\bfone-\bfr_{35}\bfrprime_{13}\right)}. \label{eq:totalDeriv}
\end{aligned}
\end{equation}
Expression \eqref{eq:totalDeriv} is of particular use in the application of the theory to the calculation of interlayer exchange coupling.

\section{Interlayer Exchange Coupling}\label{sec:iec}
We shall now proceed to describe and apply the torque method of calculating IEC. This approach is based on the observation that torque exerted on SM is due to the total out-of-plane spin current absorbed at the interface.
If we consider the thermodynamic potential $\Omega$as a function of the polarization angle $\theta$ the P and AP configurations will correspond to the values $\Omega(0)$ and $\Omega(\pi)$, respectively. We can therefore express the exchange energy $U$ as follows
\begin{equation}
U=\Omega(0)-\Omega(\pi)=-\int_{0}^{\pi}\frac{\mathrm{d}\Omega}{\mathrm{d}\theta}\mathrm{d}\theta. \label{eq:torque}
\end{equation}
Here the integrand is precisely the torque exerted on one magnetic moment by the other. 
As argued in \cite{Edwards2002}, the torque on the switching magnet is determined by the total rate of change of the out-of-plane angular momentum absorbed by it which, by continuity, equates to the net spin current absorbed by the magnet. It is therefore found as the difference between the total spin current in the spacer $J_{\nu}^{(\text{S})}$ and the right lead $J_{\nu}^{(\text{L}_2)}$: 
\begin{equation}
\frac{\mathrm{d}\Omega}{\mathrm{d}\theta}=J_{\nu}^{(\text{S})}-J_{\nu}^{(\text{L}_2)}. \label{eq:torque_current}
\end{equation}
We will suppress label $(\text{S})$ for the spacer spin current in what follows for brevity. Unlike transport current that is carried by electrons near the Fermi level $\eferm$, exchange current is contributed to by carriers at all energies from the bottom of the band and up to $\eferm$, and at all possible values of in-plane momentum $\kpar$. This leads to the following expression for the total exchange spin current
\begin{equation*}
	\begin{aligned}
		J_{\nu} &= \int_{\text{BZ}_1}\mathrm{d}\kpar\int_{-\infty}^{+\infty}\mathrm{d}E \,D(E) \\
		&\times \left(f_\text{L} \orj_\nu(\kpar,E) + f_\text{R} \olj_\nu(\kpar,E)\right),
	\end{aligned}
\end{equation*}
where
\begin{equation*}
D(E)=\frac{1}{\pi}\frac{\mathrm{d}k}{\mathrm{d}E}=\frac{1}{2\pi k}
\end{equation*}
is the density of states in the lead per unit length and spin channel, and the integration in $\kpar$ is carried out over the first Brillouin zone. The Fermi functions $f_\text{L}\equiv f(E-\mu_\text{L})$ and $f_\text{R}\equiv f(E-\mu_\text{R})$ characterise electron distributions of the left and right reservoirs with chemical potentials $\mu_\text{L}$ and $\mu_\text{R}$, respectively. In \cite{fadeev2019application} it was shown that $\olj_{\nu}=-\orj_{\nu}$ for both in-plane components $\nu=x, z$. Hence, when the system is in equilibrium, $\mu_\text{L}=\mu_\text{R}=\mu$, only the out-of-plane component ($\nu=y$) survives. Furthermore, since there can be no out-of-plane spin current in the leads, we find that \eqref{eq:torque_current} reduces to the following
\begin{equation}
\frac{\mathrm{d}\Omega}{\mathrm{d}\theta}=J_{y}=\int_{\text{BZ}_1}\mathrm{d}\kpar\int_{-\infty}^{+\infty}\mathrm{d}E \, D(E) f(E-\mu)\left(\orj_{y}+\olj_{y}\right).\label{eq:total_spincurrent_polar}
\end{equation}
From (\ref{eq:total_spincurrent_polar}) and (\ref{eq:totalDeriv}), after inserting a factor of $\tfrac{1}{2}$, since we are considering the torque transferred to one of the magnets only, we then obtain the following form of the total out-of-plane spin current
\begin{equation}
\begin{aligned}
J_y&=\frac{1}{\pi}\Im \int_{\text{BZ}_1} \op{d}\kpar \int_{-\infty}^{+\infty}\mathrm{d}E\,   f\left(E-\mu\right)\\
&\times\diff{\theta}\ln\det\bigl(\bfone-\bfr_{35}\bfrprime_{13}\bigr). \label{eq:spin_current_torque}
\end{aligned}
\end{equation}
Using \eqref{eq:torque} and \eqref{eq:torque_current} we obtain the exchange energy $U$ upon integrating over $\theta$
\begin{equation}
U = \frac{1}{\pi}\Im\int_{\text{BZ}_1}\mathrm{d}\kpar\int_{-\infty}^{+\infty}\mathrm{d}E\,f(E-\mu)\,\Bigl.\tr\ln{\left(\bfone - \bfr'_{13}(\theta)\bfr_{35}\right)}\Bigr|_0^{\pi}. \label{eq:exch_energy_torque}
\end{equation}
\section{Application of the Landauer method to a closed system}\label{sec:bc}
In this section we show that by choosing the appropriate boundary conditions we can apply the Landauer-B\"uttiker method to a closed system.
Consider a system consisting of a conducting spacer embedded in between the infinitely high insulating barriers, as shown in Figure~\ref{fig:closed-system}. We will drop the overhead arrows indicating the spatial direction on the amplitudes because there will be no ambiguity with respect to the direction of the flow in this example. There will be no electrons coming in from infinity on either side ($\bfa_1=\bfb_3=\bfzer$). Instead there will be evanescent states present in the semi-infinite layers, and propagating states in the conductor whose amplitudes are related as follows
\begin{equation}
\begin{split}
\bfa_{2}&=\bfrprime_{12}\bfb_{2},  \\
\bfb_{2}&=\bft^{-1}_{12}\bfb_{1},  
\end{split}
\qquad
\begin{split}
\bfa_{2}&=\bftprime^{-1}_{23}\bfa_{3}, \\
\bfb_{2}&=\bfr_{23}\bfa_{2},
\end{split} \label{eq:closed_amplitudes}
\end{equation}
Eliminating $\bfa_2$ and $\bfb_2$ from \eqref{eq:closed_amplitudes} we obtain
\begin{equation*}
\left(\bfone-\bfrprime_{12}\bfr_{23}\right)\bftprime^{-1}_{23}\bfa_{3} = \bfzer.
\end{equation*}
The sufficient condition for non-trivial solutions in the spacer to exist is therefore given by the equation
\begin{equation}
\det\left(\bfone - \bfrprime_{12}\bfr_{23}\right) = 0, \label{eq:det}
\end{equation}
together with the requirement that $\bftprime^{-1}_{23}\bfa_{3}$ belongs to the null-space of $\bfone-\bfr_{23}\bfrprime_{12}$. Now $(\ref{eq:det})$ is an equation in energy. Its real solutions must correspond to the permitted energies of the states in the conducting spacer, that is the energy eigenvalues of the system. The result \eqref{eq:det} readily generalises to the $N$-layer case if more layers are present between the spacer and the barriers, in which case we, as usual, label the spacer with index $n$. We will now show how the existence of these solutions is consistent with the calculation of the exchange energy in the spacer performed in earlier sections. \par
For the sake of clarity, we will carry out the derivation at a single point in momentum space and consider exchange energy density denoted $u$ given by the integral over energy in \eqref{eq:exch_energy_torque}. Integrating by parts we obtain
\begin{equation}
u=\frac{1}{\pi}\Im\left. \int_{-\infty}^{+\infty}\mathrm{d}E\, F(E) \diff{E}\ln\det\bigl(\bfone-\bfrprime_{13}(\theta)\bfr_{35}\bigr)\right|^{\theta=\pi}_{\theta=0}, \label{eq:exch_energy_ibp}
\end{equation}
where $F(E)=-k_{\text{B}}T\ln{\left[1+\exp\left(\tfrac{\mu-E}{k_{\text{B}}T}\right)\right]}$ is the anti-derivative of the Fermi distribution function. The boundary terms vanish at $E=+\infty$ suppressed by the factor of $F(E)$, and at $E=-\infty$ due to the cut off at the edge of the conducting band. 
Now defining 
\begin{equation*}
w(E,\theta) = \det\bigl(\bfone-\bfr_{35}(E,\theta)\bfrprime_{13}(E)\bigr),
\end{equation*}
we note that \eqref{eq:exch_energy_ibp} has the following form
\begin{equation}
u=\frac{1}{\pi}\Im\left.\int_{-\infty}^{+\infty}\mathrm{d}E\, F(E) \frac{\diff{E}{w(E,\theta)}}{w(E,\theta)}\right|^{\theta=\pi}_{\theta=0}. \label{eq:exch_energy_res}
\end{equation}
Integrals of the type \eqref{eq:exch_energy_res} can be evaluated in terms of the logarithmic residues of $w$ \cite{ahlfors1953complex}. Going over to the complex plane and choosing the contour $\mathcal{C}$ as shown in Figure we find \ref{fig:contour}
\figscale{0.9}{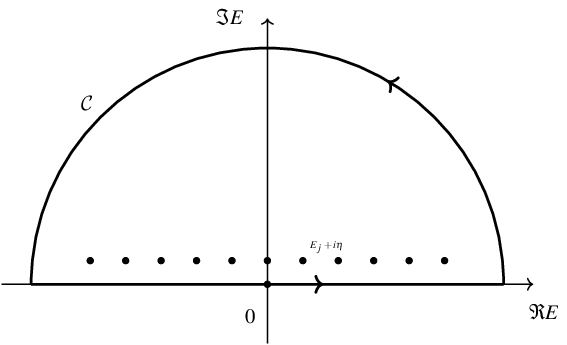}{\footnotesize Integration contour going along a segment of the real axis and closed by a semi-circular arc in the upper half plane. The dots show the positions of the roots $E_j+i\eta$ of $w(E)$, displaced by a positive infinitesimal imaginary part.}{contour}
\begin{equation}
\int_{\mathcal{C}}F(z)\frac{w'(z)}{w(z)}\mathrm{d}z = 2\pi i\sum_{j}n(\mathcal{C}, E_j)F(E_j), \label{eq:arg_principle}
\end{equation}
where $E_j$ are the roots of $w$ (the discrete energy eigenvalues) and $n(\mathcal{C}, E_j)$ are the winding numbers of $w$ with respect to $\mathcal{C}$ at $E_j$. At zero temperature
\begin{equation*}
\lim_{T\to 0_{+}} F(E) = 
\begin{cases}
E-\eferm, & \text{for}\, E < \eferm, \\
0,        & \text{for}\, E \ge \eferm,
\end{cases}
\end{equation*}
and, in the simplest case when all roots are non-degenerate, we obtain
\begin{equation}
u = 2\left(\sum_{E_j<\eferm}E_j(0)-\sum_{E_j<\eferm}E_j(\pi)\right),
\end{equation}
In other words, exchange energy in a closed system expressed in terms of the sum over the roots of equation \eqref{eq:det} is none other than the difference in energy between the parallel and anti-parallel configurations. This confirms the equivalence of the spin current (torque) and energy approaches.
\figscale{0.6}{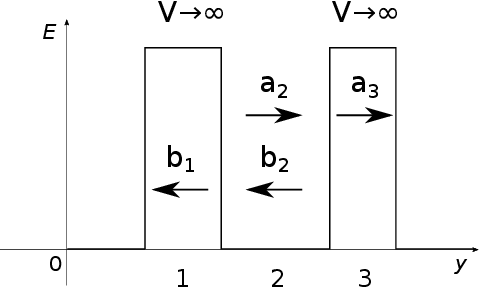}{Model of a closed system.}{closed-system}
\section{Numerical studies of the out of plane spin current}\label{sec:num}
We now explore numerically the transition process where a multilayer is gradually isolated from the leads.
  \begin{figure}%
    \centering%
    \tikzset{every picture/.style={scale=0.4, transform shape}}%
\tikzstyle{inc}=[->,shorten <=1pt, >=stealth', semithick] 
\tikzstyle{ref}=[<-,shorten <=1pt, >=stealth', semithick] 
\begin{center} 
\begin{tabular}{c}
\begin{tikzpicture} 
\draw[->](-2.5,0)--(10,0)node[below right, yshift=-0.2cm, xshift=-1cm]{\Huge{$y$}}coordinate(x axis);  
\draw[->](0,-2.5)--(0,3.5)node[above left]{\Huge{$E$}}coordinate(y axis);  
\draw[very thick](-2.5,0)--(0,0)node[below left, yshift=-0.2cm]{\Huge{$O$}};  
\draw[very thick](8.5,0)--(10,0);  
\draw[dashed](-2.5,1.3)--(10,1.3)node[above right, xshift=-1cm]{\Huge{$\eferm$}};  
\draw[very thick](0,0)--(0.5,0)--(1,0)--(1,-2)--(3,-2)--(3,0)--(6.5,0)--(6.5,-2)--(8,-2)--(8,0)--(8.5,0);  
\draw[dashed](1,-1.6)--(3,-1.6);  
\draw[dashed](6.5,-1.6)--(8.0,-1.6);  
\end{tikzpicture} \\[10pt]
\begin{tikzpicture} 
\draw[->](-2.5,0)--(10,0)node[below right, yshift=-0.2cm, xshift=-1cm]{\Huge{$y$}}coordinate(x axis);  
\draw[->](0,-2.5)--(0,3.5)node[above left]{\Huge{$E$}}coordinate(y axis);  
\draw[very thick](-2.5,0)--(0,0)node[below left, yshift=-0.2cm]{\Huge{$O$}};  
\draw[very thick](8.5,0)--(10,0);  
\draw[dashed](-2.5,1.3)--(10,1.3)node (ef) [above right, xshift=-1cm]{\Huge{$\eferm$}};  
\draw[very thick](0,0)--(0.5,0)--(0.5,1.2)--(1,1.2)--(1,-2)--(3,-2)--(3,0)--(6.5,0)--(6.5,-2)--(8,-2)--(8,1.2)--(8.5,1.2)--(8.5,0) ;  
\draw[dashed](1,-1.6)--(3,-1.6);  
\draw[dashed](6.5,-1.6)--(8.0,-1.6);  
\end{tikzpicture} \\[10pt]
\begin{tikzpicture} 
\draw[->](-2.5,0)--(10,0)node[below right, yshift=-0.2cm, xshift=-1cm]{\Huge{$y$}}coordinate(x axis);  
\draw[->](0,-2.5)--(0,3.5)node[above left]{\Huge{$E$}}coordinate(y axis);  
\draw[very thick](-2.5,0)--(0,0)node[below left, yshift=-0.2cm]{\Huge{$O$}};  
\draw[very thick](8.5,0)--(10,0);  
\draw[dashed](-2.5,1.3)--(10,1.3)node[above right, xshift=-1cm]{\Huge{$\eferm$}};  
\draw[very thick](0,0)--(0.5,0)--(0.5,4)--(1,4)node[above,midway]{}--(1,-2)--(3,-2)--(3,0)--(6.5,0)--(6.5,-2)--(8,-2)--(8,4)--(8.5,4)node[above,midway]{}--(8.5,0) ;  
\draw[dashed](1,-1.6)--(3,-1.6);  
\draw[dashed](6.5,-1.6)--(8.0,-1.6);  
\end{tikzpicture} 
\end{tabular}
\end{center}
\caption{\footnotesize CPP multi-layer with additional barriers gradually isolating it from the leads.}
\label{fig:potential_profiles}%
  \end{figure}
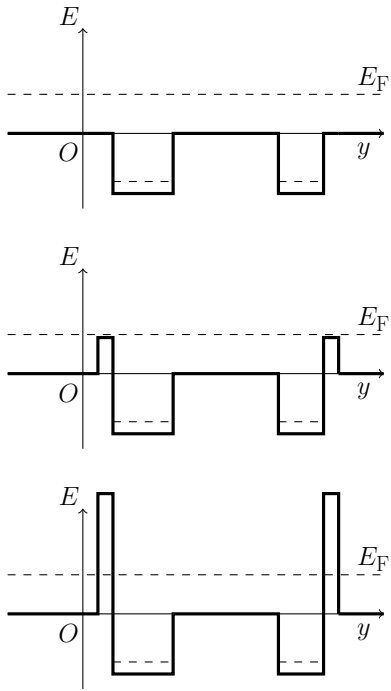%

\begin{figure}[tp]
	\begin{center}
		\includegraphics[width=1.0\linewidth]{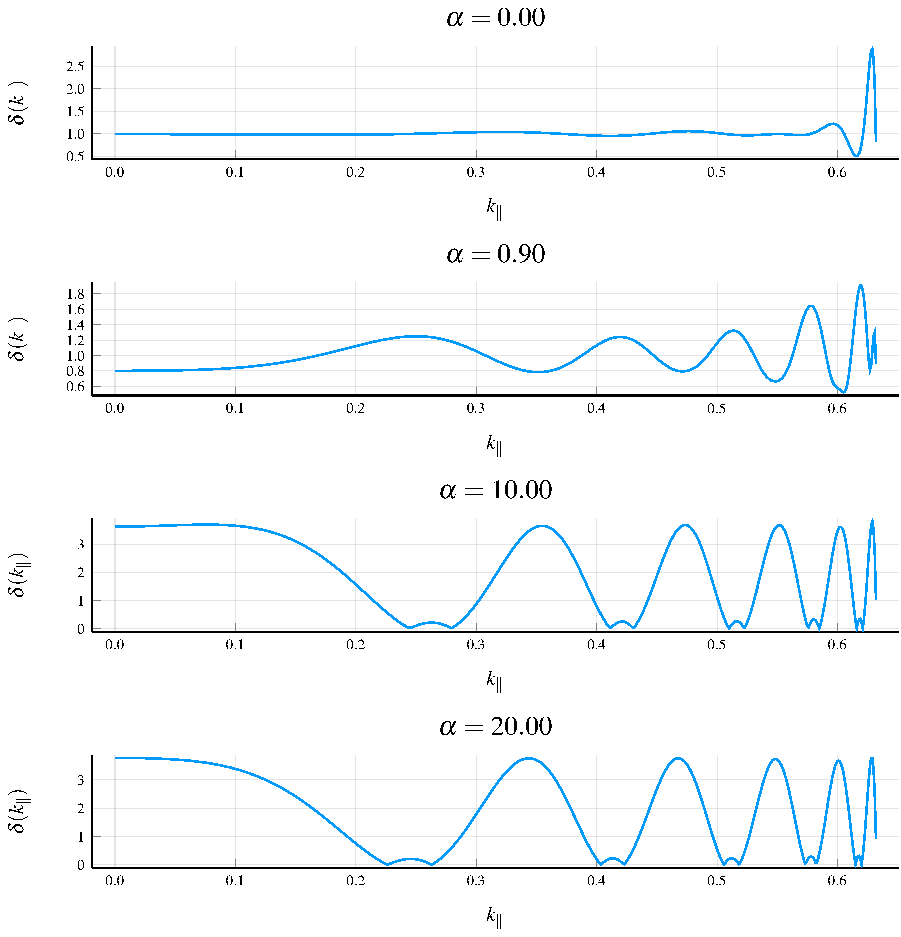}
	\end{center}
	\caption{\footnotesize $\delta(\kpar)$ plotted for different values of $\alpha=V/\eferm$. In the closed regime ($\alpha>1$) distinct roots occur in momentum space.}
	\label{fig:det_roots_k_space}
\end{figure}

We start with the same 5-layer CPP structure considered in the previous chapter (see \Figref{fig:cpp}). We then insert a pair of extra barriers between the leads and the magnets, as shown in \Figref{fig:potential_profiles}, with the parameters given in Table~\ref{tbl:extra-barriers}.
\begin{table}[ht]
	\centering
	\begin{tabular}{|c|c|c|c|c|}
		\hline 
		& $V$ & $\Delta$ & $\theta$ & $y_{n+1}-y_{n}$\tabularnewline
		\hline 
		$L_1$ & 0.0 & 0.0 & 0.0 & -\tabularnewline
		\hline 
		$B_1$ & $\alpha\eferm$ & 0.0 & 0.0 & 1.0\tabularnewline
		\hline
		PM & -0.1 & 0.05 & 0.6 & 7.0\tabularnewline
		\hline 
		S & 0.0 & 0.0 & 0.0 & 20.0\tabularnewline
		\hline 
		SM & -0.1 & 0.05 & 0.0 & 3.0\tabularnewline
		\hline 
		$B_2$ & $\alpha\eferm$ & 0.0 & 0.0 & 1.0\tabularnewline
		\hline
		$L_2$ & 0.0 & 0.0 & 0.0 & -\tabularnewline
		\hline
	\end{tabular}
	\caption{\footnotesize Parameters of the model used to demonstrate the process of gradually turning an open system into a closed one. Extra potential barriers $B_1$ and $B_2$ are added between the leads and the magnets. The barrier height is then increased, which is controlled by parameter $\alpha=\frac{V}{E_{\text{F}}},$.}
	\label{tbl:extra-barriers}
\end{table}
We show four stages of the transition process by varying the dimensionless parameter
\begin{equation*}
\alpha=\frac{V}{E_{\text{F}}},
\end{equation*}
where $V$ is the potential barrier height, and the energy scale is chosen so that $\eferm > 0$. In the first configuration $\alpha=0$, that is the system is open, free from the influence of the extra barriers. In the second one, $0<\alpha<1$, the height of the barriers is set between the level of the leads/spacer and the Fermi energy level,  the system is partially confined. Further, the barrier height is set above the Fermi level, $\alpha>1$, and the system is isolated. Lastly, in order to demonstrate the tendency in the current when the confinement is further increased, we produce the plot with $\alpha\gg 1$, where the system is strongly confined.
In order to show the emergence of bound states in the spacer we plot the following function (note that with the addition of the barriers we now have a 7-layer system, that is $N=7$, and the spacer index is $n=4$)
\begin{equation*}
\delta(\kpar)=\left|\det\left(\bfone-\bfr^{\phantom{'}}_{47}\bfr'_{14}\right)\right|,
\end{equation*}
as shown in Figure~\ref{fig:det_roots_k_space}. We take the absolute value merely for illustrative purposes, which does not alter the positions of the roots. We observe that as the barrier height is increased $\delta(\kpar)$ develops a sequence of  roots at isolated values of $\kpar$. These correspond to the positions of energy eigenvalues of the closed system at $\eferm$. 

\begin{figure}[tp]
	\begin{center}
		\includegraphics[width=1.0\linewidth]{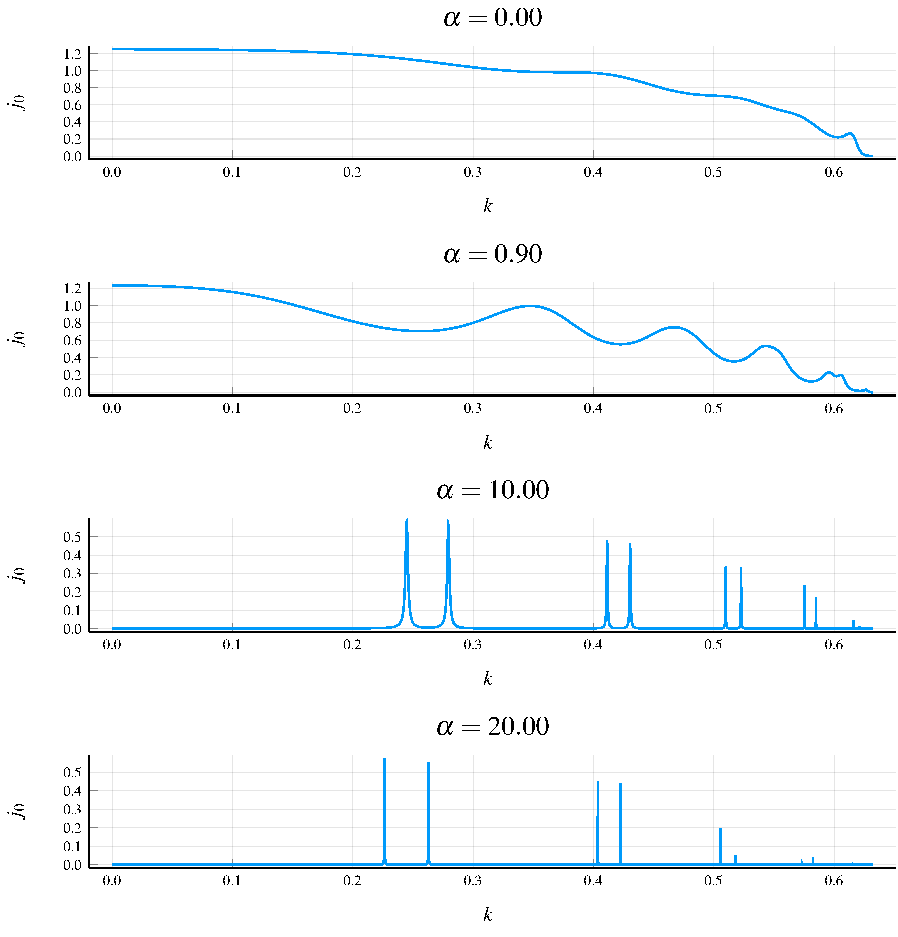}
	\end{center}
	\caption{\footnotesize Charge current density in momentum space, plotted at different values of $\alpha=V/\eferm$, as the system is gradually turned to a closed one.}
	\label{fig:k_space_closed_0}
\end{figure}

\begin{figure}[tp]
	\begin{center}
		\includegraphics[width=1.0\linewidth]{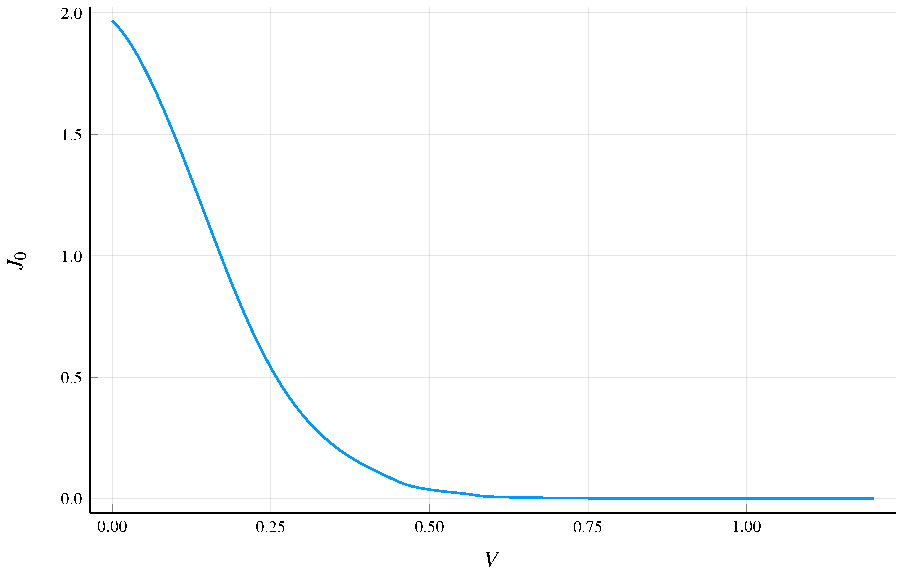}
	\end{center}
	\caption{\footnotesize Charge current integrated over in-plane momentum, plotted as a function of the increasing insulating potential barrier height.}
	\label{fig:current_barriers_0}
\end{figure}

\begin{figure}[tp]
	\begin{center}
		\includegraphics[width=1.0\linewidth]{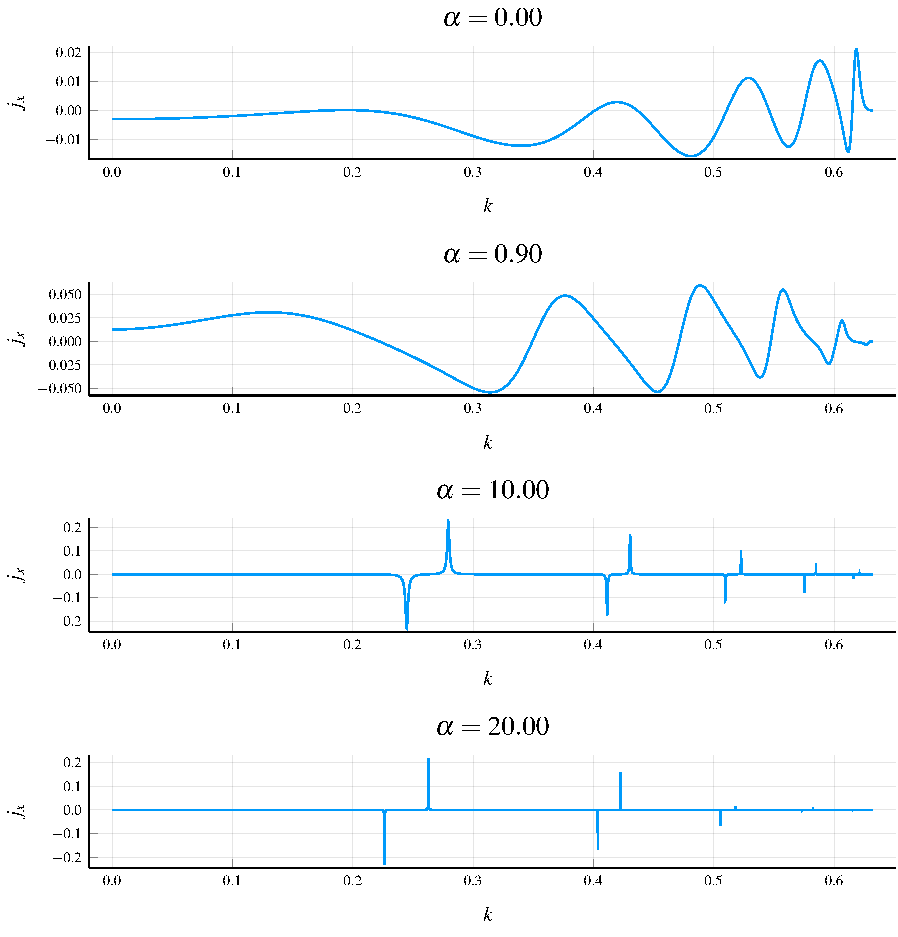}
	\end{center}
	\caption{\footnotesize In-plane spin current density in momentum space, plotted at different values of $\alpha=V/\eferm$, as the system is gradually turned to a closed one.}
	\label{fig:k_space_closed_x}
\end{figure}

\begin{figure}[tp]
	\begin{center}
		\includegraphics[width=1.0\linewidth]{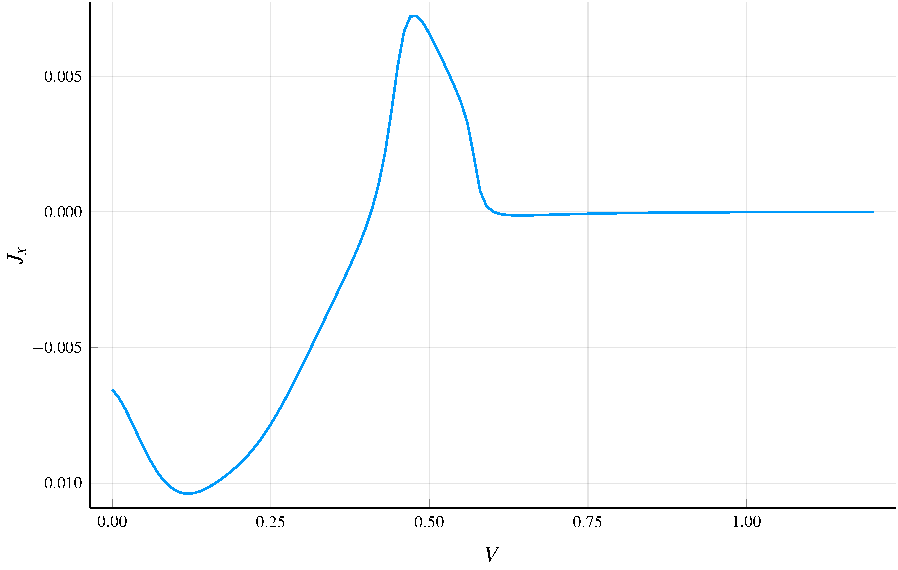}
	\end{center}
	\caption{\footnotesize In-plane spin current integrated over in-plane momentum, plotted as a function of the increasing insulating potential barrier height.}
	\label{fig:current_barriers_x}
\end{figure}

\begin{figure}[tp]
	\begin{center}
		\includegraphics[width=1.0\linewidth]{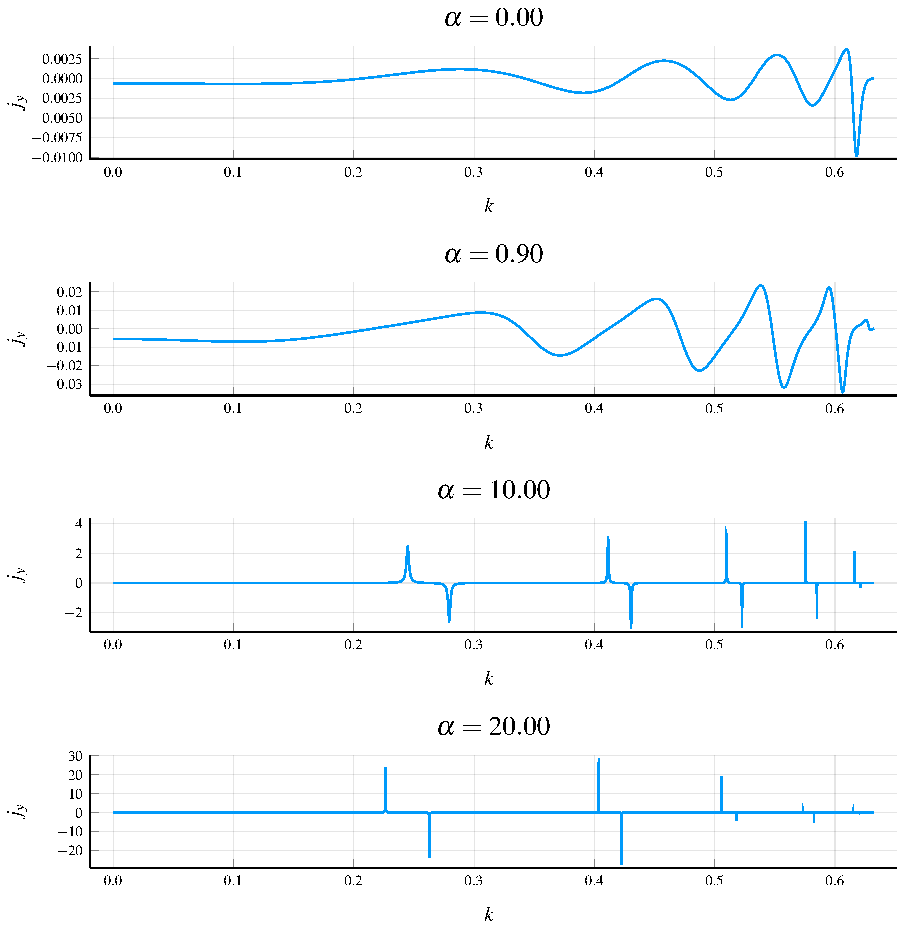}
	\end{center}
	\caption{\footnotesize Out-of-plane spin current density in momentum space, plotted at different values of $\alpha=V/\eferm$, as the system is gradually turned to a closed one.}
	\label{fig:k_space_closed_y}
\end{figure}

\begin{figure}[tp]
	\begin{center}
		\includegraphics[width=1.0\linewidth]{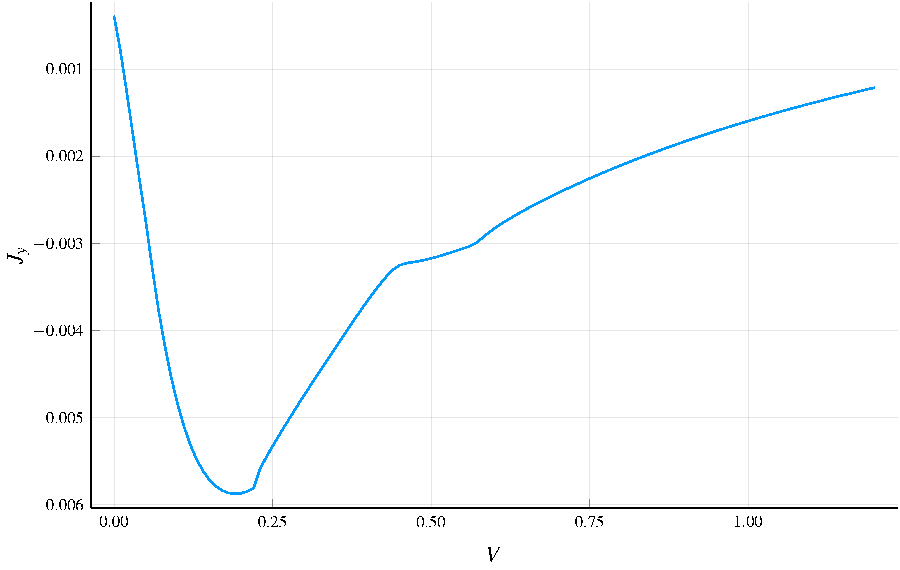}
	\end{center}
	\caption{\footnotesize Out-of-plane spin current integrated over in-plane momentum, plotted as a function of the increasing insulating potential barrier height.}
	\label{fig:current_barriers_y}
\end{figure}

\begin{figure}[tp]
	\begin{center}
		\includegraphics[width=1.0\linewidth]{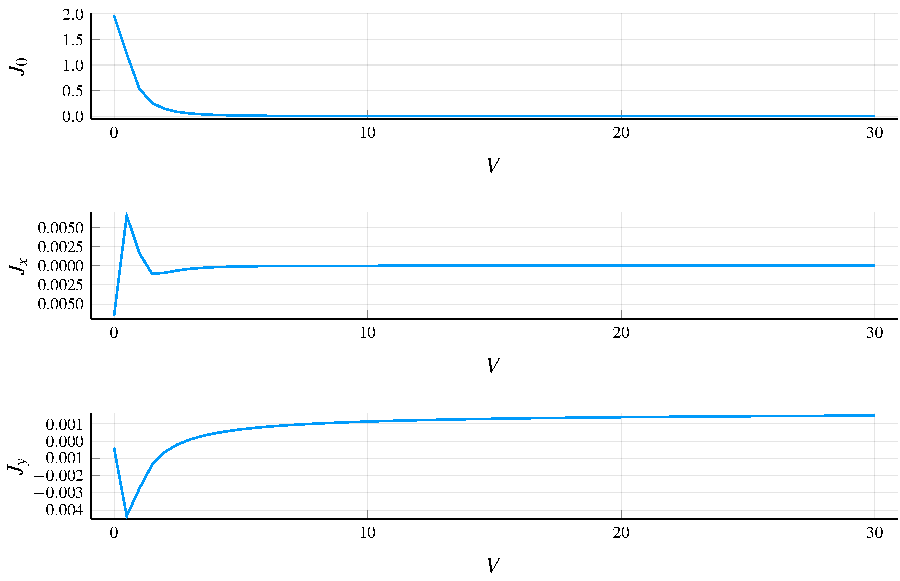}
	\end{center}
	\caption{\footnotesize Current integrated over in-plane momentum, plotted as a function of the increasing insulating potential barrier height. Extended range shows the slow divergence if the out-of plane component, contrasted against the rapid vanishing of the other components.}
	\label{fig:current_barriers_combined}
\end{figure}

We next look at the behaviour of the three spin current components in momentum space. In Figure~\ref{fig:k_space_closed_0} we see that the charge current density $j_0$ develops a series of sharp resonances that precisely correspond to the positions of the bound states of the system. The height of these resonances gradually decreases with stronger confinement. In agreement with that behaviour, the total current $J_0$ (integrated over the in-plane momentum) vanishes with the increasing barrier height, Figure~\ref{fig:current_barriers_0}. This is fully expected because charge current cannot flow in an isolated system.%
Next, in Figure~\ref{fig:k_space_closed_x} we show the plot of the in-plane spin current density $j_x$. Similarly to the charge component, we observe resonances in $\kpar$-space at the positions of the bound states that decay with the increase of the barrier height. The total in-plane spin current $J_x$ also vanishes quickly after an initial increase, as shown in Figure~\ref{fig:current_barriers_x}.%
Lastly, in Figure~\ref{fig:k_space_closed_y} we display the behaviour of the out-of-plane component $j_y$. Here the situation is qualitatively different, as the resonances provide non-vanishing contributions under increasing confinement, and the total current therefore does not converge to zero, as seen in Figure~\ref{fig:current_barriers_y}. This is consistent with the fact that IEC survives in a closed system.%
In order to illustrate the point further, we plot the integrated values of all three components against the increasing barrier height, Figure~\ref{fig:current_barriers_combined}. It is clearly seen that the total out-of-plane diverges, while the other two components disappear, as the system is isolated.
\section{Conclusion}
We have demonstrated here, through an analytical argument and a numerical example, that the application of the Landauer-B\"uttiker formalism to the study of ballistic transport yields consistent results whether the system under consideration is open, i.e. is supplied with electrons at all possible momenta from macroscopic reservoirs, or closed, i.e. sealed off of the environment by infinite potential barriers. We have also stated the boundary conditions applicable in the case of a closed system. Mathematically, this consistency is achieved, on the one hand, through evaluation of a specific integral via a series of delta terms, and on the other hand, by summing over the allowed discrete states arising in a closed system. In this process the out-of-plane spin current component represents a special case because it happens to be an exact derivative with respect to the polarisation angle $\theta$ that the magnetic moment of the PM makes with the vertical axis. From the physical point of view the out-of-plane current exerts a torque on the SM which is related to the IEC energy. As the system is gradually isolated from the leads by high barrier potentials this current collapses into a set of resonances, and can therefore be evaluated by summing over those. As the barrier potentials are increased conducting electrons in the spacer become fully reflected from both barriers, which leads to the boundary conditions and the eigenvalue equation (\ref{eq:det}).
\section{Appendix A: Proof of Equation~(\ref{eq:totalDeriv})} \label{app:A}
We shall now prove that the out-of-plane current density is an exact derivative in the polarisation angle. This fact is of crucial importance in establishing the relation between torque and energy methods of calculating IEC. We sum over the left- and right-moving parts of both the left- and right-incident current in the layer of interest. We derive the general result for an $n$-th segment of an $N$-layer system which is valid as long as the $n$-th layer is non-magnetic. For simplicity we assume that only the magnetization of the layers adjacent to the spacer is set at an angle of $\theta$ in-plane.

Equations (\ref{eq:orab}) and the corresponding results for the left-moving current generalise in a straightforward way to an $N$-layer device:
\begin{equation}
\begin{aligned}
\overrightarrow{\bfa} & =\left(\bfI-\bfr^{'}_{1n}\bfr_{nN}\right)^{-1}\bft^{'}_{1n} \\ \overrightarrow{\bfb} & =\bfr_{nN}\overrightarrow{\bfa}=\bfr_{nN}\left(\bfI-\bfr^{'}_{1n}\bfr_{nN}\right)^{-1}\bft^{'}_{1n}\\
\overleftarrow{\bfb} & =\left(\bfI-\bfr_{nN}\bfr^{'}_{1n}\right)^{-1}\bft_{nN} \\ \overleftarrow{\bfa} & =\bfr^{'}_{1n}\overleftarrow{\bfb}=\bfr^{'}_{1n}\left(\bfI-\bfr_{nN}\bfr^{'}_{1n}\right)^{-1}\bft_{nN}
\end{aligned}
\end{equation}
For brevity we introduce additional notation for the following factors
\begin{equation}
  \begin{aligned}
    \overrightarrow{\bfr}_n &= \left(\bfI-\bfr^{'}_{1n}\bfr_{nN}\right)^{-1} \\
    \overleftarrow{\bfr}_n &= \left(\bfI-\bfr_{nN}\bfr^{'}_{1n}\right)^{-1} \\
  \end{aligned} \label{eq:ref_count_exp_n}
\end{equation}
which have the physical meaning of the total right- and left-reflected amplitudes in layer $n$ accounting for all reflections within the intermediate layers between $n$ and the leads (1 and $N$). For what follows it is useful to note the following relations which can be easily verified by expanding the geometric series:
\begin{equation}
\begin{aligned}
\bfr_{nN}\overrightarrow{\bfr}_n &= \overleftarrow{\bfr}_n\bfr_{nN} \\
\overrightarrow{\bfr}_n\bfr^{'}_{1n} &= \bfr^{'}_{1n}\overleftarrow{\bfr}_n
\end{aligned} \label{eq:reciprocal}
\end{equation}
We shall now prove that the out-of-plane current density is an exact derivative with respect to the polarisation angle. Since we are interested in the exchange part of the current we must add the left- and right-moving parts of both the left- and right-incident current in the layer of interest. We derive the general result for an $n$-th (non-magnetic) segment of an $N$-layer system. For simplicity we assume that only the magnetisation of the layer adjacent to the spacer on the left is set at an angle of $\theta_{n-1} = \theta$ in-plane, and $\theta_k=0,\, k\ne n-1$. Calculating the out-of-plane spin current due to electrons emerging from the left reservoir we obtain
\begin{equation}
\begin{aligned}
\frac{1}{k_n}\orj_y &= \tr{\left\{\orbA^{\dag}\bfsig_{y}\orbA - \orbB^{\dag}\bfsig_y\orbB\right\}} \\
&=\tr{\left\{\bftprime_{1n}^{\dag}\overrightarrow{\bfr}^{\dag}_n\bfsig_y\overrightarrow{\bfr}_n^{\phantom{\dag}}\bftprime^{\phantom{\dag}}_{1n}\right\}}\\
&\phantom{=}-\tr\left\{\bftprime_{1n}^{\dag}\overrightarrow{\bfr}^{\dag}_n\bfr_{nN}^{\dag}\bfsig_y \bfr_{nN}^{\phantom{\dag}}\overrightarrow{\bfr}_n^{\phantom{\dag}}\bftprime^{\phantom{\dag}}_{1n}\right\},
\end{aligned}
\end{equation}
where $\overrightarrow{\bfr}_n,\,\overleftarrow{\bfr}_n$ are defined as in \eqref{eq:ref_count_exp_n}. Using the cyclic property of trace and the relations 
\begin{subequations}
	\begin{align*}
	\overrightarrow{\bfr}_{n}\bfrprime_{1n} &=\bfrprime_{1n}\overleftarrow{\bfr}_{n},\\
	\bfr_{nN}\overrightarrow{\bfr}_{n} &=\overleftarrow{\bfr}_{n}\bfr_{nN},
	\end{align*} 
\end{subequations}
we can transform the right-hand side as follows
\begin{equation}
\begin{aligned}
\frac{1}{k_n}\orj_y &= \tr\left\{\overrightarrow{\bfr}_n^{\phantom{\dag}}\bftprime^{\phantom{\dag}}_{1n}\bftprime^{\dag}_{1n}\overrightarrow{\bfr}^{\dag}_n\bfsig_y\right\}\\ &\phantom{=}-\tr\left\{\overleftarrow{\bfr}_n^{\phantom{\dag}}\bfr_{nN}^{\phantom{\dag}}\bftprime^{\phantom{\dag}}_{1n}\bftprime^{\dag}_{1n}\bfr^{\dag}_{nN}\overleftarrow{\bfr}^{\dag}_n \bfsig_y\right\}.
\label{eq:angle_deriv_proof_right}
\end{aligned}\end{equation}
We can now use conservation of charge current to eliminate transmission matrices from (\ref{eq:angle_deriv_proof_right}) and work only with reflection coefficients thereafter. Taking advantage of the fact that the spacer and the leads are non-magnetic we obtain
\begin{equation*}
\bftprime^{\phantom{\dag}}_{1n}\bftprime^{\dag}_{1n} = \bfk^{\phantom{1}}_1\bfk_n^{-1}\left(\bfone - \bfrprime^{\phantom{\dag}}_{1n}\bfrprime^{\dag}_{1n}\right).
\end{equation*}
With that in mind, (\ref{eq:angle_deriv_proof_right}) is written as follows
\begin{equation}
\begin{aligned}
\frac{1}{k_1}\orj_y &= \tr\left\{\overrightarrow{\bfr}_n^{\phantom{\dag}}\overrightarrow{\bfr}^{\dag}_n\bfsig_y\right\} - \tr\left\{\overrightarrow{\bfr}_n^{\phantom{\dag}}\bfrprime^{\phantom{\dag}}_{1n}\bfrprime^{\dag}_{1n}\overrightarrow{\bfr}^{\dag}_n\bfsig_y\right\} \\
&\phantom{=}-\tr\left\{\overleftarrow{\bfr}_n^{\phantom{\dag}}\bfr^{\phantom{\dag}}_{nN}\bfr^{\dag}_{nN}\overleftarrow{\bfr}^{\dag}_n \bfsig_y\right\}\\ &\phantom{=}+\tr\left\{\overleftarrow{\bfr}_n^{\phantom{\dag}}\bfr^{\phantom{\dag}}_{nN}\bfrprime^{\phantom{\dag}}_{1n}\bfrprime^{\dag}_{1n}\bfr^{\dag}_{nN}\overleftarrow{\bfr}^{\dag}_n \bfsig_y\right\}.
\end{aligned} \label{eq:proof_right}
\end{equation}
Following similar steps for the left-moving current and noting that
\begin{equation*}
\bft^{\phantom{\dagger}}_{nN}\bft^{\dagger}_{nN} = \bfk^{\phantom{1}}_N\bfk_n^{-1}\left(\bfone - \bfr^{\phantom{\dagger}}_{nN}\bfr^{\dagger}_{nN}\right),
\end{equation*}
we obtain
\begin{equation}
\begin{aligned}
\frac{1}{k_N}\olj_y &=
\tr\left\{\overrightarrow{\bfr}_n^{\phantom{\dagger}}\bfrprime^{\phantom{\dagger}}_{1n}\bfrprime^{\dagger}_{1n}\overrightarrow{\bfr}^{\dagger}_n\bfsig_y\right\}\\ &\phantom{=}-\tr\left\{\overrightarrow{\bfr}_n^{\phantom{\dagger}}\bfrprime^{\phantom{\dagger}}_{1n}\bfr^{\phantom{\dagger}}_{nN}\bfr^{\dagger}_{nN}\bfrprime^{\dagger}_{1n}\overrightarrow{\bfr}^{\dagger}_n \bfsig_y\right\} \\
&\phantom{=}- \tr\left\{\overleftarrow{\bfr}_n^{\phantom{\dagger}}\overleftarrow{\bfr}^{\dagger}_n\bfsig_y\right\} +\tr\left\{\overleftarrow{\bfr}_n^{\phantom{\dagger}}\bfr^{\phantom{\dagger}}_{nN}\bfr^{\dagger}_{nN}\overleftarrow{\bfr}^{\dagger}_n \bfsig_y\right\}.
\end{aligned} \label{eq:proof_left}
\end{equation}
Since we assume $k_1 = k_N=k$ throughout, adding (\ref{eq:proof_right}) and (\ref{eq:proof_left}) we can calculate the total exchange current density
\begin{equation}
\begin{aligned}
\frac{1}{k}j_y &= \frac{1}{k}\left(\orj_y+\olj_y\right) \\
&=\tr\left\{\overrightarrow{\bfr}^{\phantom{\dagger}}_n\left(\bfone-\bfrprime^{\phantom{\dagger}}_{1n}\bfr^{\phantom{\dagger}}_{nN}\bfr^{\dagger}_{nN}\bfrprime^{\dagger}_{1n}\right)\overrightarrow{\bfr}^{\dagger}_n\bfsig_y\right\} \\
&\phantom{=}-\tr\left\{\overleftarrow{\bfr}^{\phantom{\dagger}}_n\left(\bfone-\bfr^{\phantom{\dagger}}_{nN}\bfrprime^{\phantom{\dagger}}_{1n}\bfrprime^{\dagger}_{1n}\bfr^{\dagger}_{nN}\right)\overleftarrow{\bfr}^{\dagger}_n\bfsig_y\right\} \\
&=\tr{\left\{ \left(\overrightarrow{\bfr}_n-\overleftarrow{\bfr}_n\right)\bfsig_y +\text{h.c.}\right\} } \\
&=\tr{\left\{ \left(\bfone-\bfr_{nN}\bfrprime_{1n}\right)^{-1}\bfr_{nN}\left[\bfsig_y,\bfrprime_{1n}\right]+\text{h.c.}\right\} }\\
&=-2i\tr{\left\{ \left(\bfone-\bfr_{nN}\bfrprime_{1n}\right)^{-1}\bfr^{\phantom{\dagger}}_{nN}\left(\frac{i}{2}\left[\bfsig_y,\bfrprime_{1n}\right]\right)-\text{h.c.}\right\}},
\end{aligned} \label{eq:angle_deriv_total}
\end{equation}
where $\text{h.c.}$ stands for the Hermitian conjugate of the preceding term and $[,]$ is the standard commutator. Only reflections from the left represented by the factor $\bfrprime_{1n}$ accrue polarisation, and the angular dependence is given by the following formula
\begin{equation}
\bfrprime_{1n}(\theta) = e^{-\frac{i\bfsig_y\theta}{2}}\bfr'_{1n}(0) e^{\frac{i\bfsig_y\theta}{2}}. \label{eq:angle_dependence}
\end{equation}
Differentiating (\ref{eq:angle_dependence}) with respect to $\theta$ we can write
\begin{equation}
\frac{\mathrm{d}}{\mathrm{d}\theta}\bfrprime_{1n}(\theta) = -\frac{i}{2}\left[\bfsig_y,\bfrprime_{1n}\right]. \label{eq:deriv_wrt_theta}
\end{equation}
Substituting (\ref{eq:deriv_wrt_theta}) into (\ref{eq:angle_deriv_total}) and using the chain rule we obtain
\begin{equation}
j_y = -2i k \frac{\mathrm{d}}{\mathrm{d}\theta}\tr\left\{\ln{\left(\bfone-\bfr_{nN}\bfrprime_{1n}\right)} - \text{h.c.}\right\}. \label{eq:angle_derivative_pre_result}
\end{equation}
Rewriting \eqref{eq:angle_derivative_pre_result} in terms of taking the imaginary part we finally arrive at the desired result
\begin{equation}
j_y = -4k \Im\frac{\mathrm{d}}{\mathrm{d}\theta}\ln\det{\left(\bfone-\bfr_{nN}\bfrprime_{1n}\right)}, \label{eq:angle_derivative_result}
\end{equation}
in the form of an exact derivative with respect to the magnetisation direction.
\section{Appendix B: calculation using the energy method}
In this section we calculate IEC using the energy method, loosely adapting the argument given in \cite{PhysRevB.52.411} where it is done for the multi-orbital case. We do not need the full generality here. Instead we make the proof compatible with the transfer matrix formalism in order to compare the result with that obtained using the torque method.

We begin by considering a system with two magnets separated by a non-magnetic conducting spacer. The essence of the energy method is in calculating the difference between the thermodynamic potentials in the magnets, expressed in terms of the local density of states. This allows one to resolve the total energy in the up- and down-spin population, which is required for computing IEC. We thus define the difference in thermodynamic potentials of the system $U$ when the magnetisations of the magnets are in the P and AP alignment, respectively
\begin{equation}
U=\Omega_{\text{P}}^{\ua}+\Omega_{\text{P}}^{\da}- \Omega_{\text{AP}}^{\ua}-\Omega_{\text{AP}}^{\da}.\label{eq:exch_int_general}
\end{equation}
Here $\Omega_{\text{P/AP}}^{\sigma}$ is the thermodynamic potential for electrons of spin orientation $\sigma$ in a system where the magnetization is the P or AP state, and at finite temperature is given by the following formula \cite{PhysRevB.52.411}:
\begin{equation}
\Omega^{\sigma}=\int_{-\infty}^{+\infty}\mathrm{d}E\,F(E)\rho^{\sigma}(E), \label{eq:grand_potential}
\end{equation}
where 
\begin{equation*}
F(E)=-k_{\text{B}}T\ln{\left[1+\exp\left(\tfrac{\mu-E}{k_{\text{B}}T}\right)\right]}
\end{equation*}
is the anti-derivative of the Fermi distribution function and $\rho^{\sigma}(E)$ is the spin-resolved local density of states in the spacer, given in terms of the one-particle Green's function:
\begin{equation}
\rho^{\sigma}(E,L) = -\frac{1}{\pi}\Im\, \tr g_\text{LR}^{\sigma}(E+i 0^{+}), \label{eq:ldos} 
\end{equation}
where $g^{\sigma}_\text{LR}$ is the part of the spacer Green's function accounting for the interaction of the magnets via conduction electrons in the spacer, and $L$ is the spacer thickness. Precisely, this means the following. Let $g_{\text{S}}$ be the Green's function in the spacer of the original multilayer. Let $g_{\text{L}}$, $g_{\text{R}}$ be the Green's functions in the spacer calculated in the presence of only one of the magnets (to the left or to the right, respectively), as if the other one did not exist. Then $g_\text{LR}$ is found from the following expansion
\begin{equation*}
g_{\text{S}} = g_0 + g_{\text{L}} + g_{\text{R}} + g_{\text{LR}},
\end{equation*}
where $g_0$ is the free-particle Green's function.
We find that the exchange energy is expressed as follows
\begin{equation}
U = \sum_{\sigma=\ua,\da}\int_{-\infty}^{+\infty}\mathrm{d}E\,F(E)\left[\rho^{\sigma}_{\text{P}}(E)-\rho^{\sigma}_{\text{AP}}(E)\right], \label{eq:exch_energy_rho}
\end{equation}
where P and AP signify that the density of states is calculated separately for each alignment of the magnets in the system, respectively. Expressing the sum over the spin orientations in terms of taking a trace we obtain
\begin{equation*}
\rho(E,l)=\sum_{\sigma=\uparrow,\downarrow}\rho^{\sigma}(E,L)=-\frac{1}{\pi}\Im\tr\bfg_\text{LR}(E+i 0^{+}),
\end{equation*}
we use the following result featured in \cite{PhysRevB.52.411} and \cite{10.1007/3-540-46437-9_9}
\begin{equation}
\tr \bfg_\text{LR}=\diff{E}\tr\ln{\left(\bfone - \bfg_0\bftau_\text{L}\bfg_0\bftau_\text{R}\right)}, \label{eq:greens_deriv_bruno}
\end{equation}
where
\begin{equation*}
\bftau_i=\bfv_i\left(\bfone - \bfg_0\bfv_i\right)^{-1},
\end{equation*}
known as the $T$-matrix (not the same as the transfer matrix $\bfT$!). Using the following relations \cite{economou2006green}
\begin{subequations} \label{eq:t-matrix}
	\begin{align}
	\bfr &= \Braket{y|\bfg_0|y}\Braket{-k|\bftau|k}, \\
	\Braket{-k|\bftau|k} &= \int e^{i(x'+x'')}\Braket{x''|\bftau|x'}dx'dx'',
	\end{align}
\end{subequations}
inserting the resolution of identity and integrating, we can express \eqref{eq:greens_deriv_bruno} in terms of the reflection matrices
\begin{equation}
\tr \bfg_\text{LR}=\diff{E}\tr\ln{\left(\bfone - \bfrprime_{13}\bfr_{35}\right)}, \label{eq:greens_deriv}
\end{equation}
which is equivalent to equation (5.16) in \cite{PhysRevB.52.411}. In this form it is clear that the P and AP alignments correspond to setting $\theta=0$ and $\theta=\pi$ in $\bfrprime_{1n}(\theta)$, respectively. Substituting \eqref{eq:greens_deriv} into \eqref{eq:exch_energy_rho} and integrating by parts we finally obtain
\begin{equation}
U = \frac{1}{\pi}\Im\int_{\text{BZ}_1}\mathrm{d}\kpar\int_{-\infty}^{+\infty}\mathrm{d}E\,f(E-\mu)\,\Bigl.\tr\ln{\left(\bfone - \bfrprime_{13}(\theta)\bfr_{35}\right)}\Bigr|_0^{\pi}, \label{eq:exch_energy}
\end{equation}
which matches exactly the result \eqref{eq:exch_energy_torque} obtained earlier using the torque method.
\bibliographystyle{unsrt}
\bibliography{bibliography}
\end{document}